\documentclass[iop]{emulateapj}
\usepackage{graphicx}
\shorttitle{DDO216-A1 Star Cluster}
\shortauthors{Cole et al.}
\begin{document}
%
%
\newcounter{Affil}
\newcommand{\Affil}{\refstepcounter{Affil}{\theAffil}}
\setcounter{Affil}{1}
\title{DDO216-A1: a Central Globular Cluster in a Low-Luminosity Transition Type Galaxy\altaffilmark{\theAffil}}
\author{Andrew A. Cole\altaffilmark{\Affil\label{Affil:utas}}, 
Daniel R. Weisz\altaffilmark{\Affil\label{Affil:berk}},
Evan D. Skillman\altaffilmark{\Affil\label{Affil:minn}}, 
Ryan Leaman\altaffilmark{\Affil\label{Affil:mpifa}},
Benjamin F. Williams\altaffilmark{\Affil\label{Affil:wash}},
Andrew E. Dolphin\altaffilmark{\Affil\label{Affil:ray}},
L. Clifton Johnson\altaffilmark{\Affil\label{Affil:ucsd}},
Alan W. McConnachie\altaffilmark{\Affil\label{Affil:hia}},
Michael Boylan-Kolchin\altaffilmark{\Affil\label{Affil:tex}},
Julianne Dalcanton\altaffilmark{\ref{Affil:wash}},
Fabio Governato\altaffilmark{\ref{Affil:wash}},
Piero Madau\altaffilmark{\Affil\label{Affil:ucsc}},
Sijing Shen\altaffilmark{\Affil\label{Affil:ioacam}},
Mark Vogelsberger\altaffilmark{\Affil\label{Affil:kavmit}}
\setcounter{Affil}{1}
}
\altaffiltext{1}{Based on observations made with the NASA/ESA Hubble Space Telesope,
obtained at the Space Telescope Science Institute, which is operated by the Association
of Universities for Research in Astronomy, Inc., under NASA contract NAS 5-26555. These
observations were obtained under program GO-13768.}
\altaffiltext{\ref{Affil:utas}}{School of Physical Sciences, University of Tasmania, Private Bag 37, Hobart, 
Tasmania, 7001 Australia; andrew.cole@utas.edu.au}
\altaffiltext{\ref{Affil:berk}}{Department of Astronomy, University of California Berkeley, Berkeley, CA 94720, USA}
\altaffiltext{\ref{Affil:minn}}{Minnesota Institute for Astrophysics, University of Minnesota, Minneapolis, MN 55441, USA; skillman@astro.umn.edu}
\altaffiltext{\ref{Affil:mpifa}}{Max-Planck Institut f\"{u}r Astronomie, K\"{o}nigstuhl 17, D-69117, Heidelberg, Germany}
\altaffiltext{\ref{Affil:wash}}{Department of Astronomy, University of Washington, Box 351580, Seattle, WA 98195 USA}
\altaffiltext{\ref{Affil:ray}}{Raytheon; 1151 E. Hermans Rd, Tucson, AZ 85756, USA; adolphin@raytheon.com}
\altaffiltext{\ref{Affil:ucsd}}{Center for Astrophysics and Space Sciences, University of California, San Diego, 9500 Gilman Drive, La Jolla, CA 92093, USA}
\altaffiltext{\ref{Affil:hia}}{National Research Council, Herzberg Institute of Astrophysics, 5071 West Saanich Road, Victoria, BC, Canada V9E~2E7; alan.mcconnachie@nrc-cnrc.gc.ca}
\altaffiltext{\ref{Affil:tex}}{Department of Astronomy, The University of Texas at Austin, 2515 Speedway, Stop C1400, Austin, TX 78712-1205, USA}
\altaffiltext{\ref{Affil:ucsc}}{Department of Astronomy and Astrophysics, University of California, 1156 High Street, Santa Cruz CA 95064, USA}
\altaffiltext{\ref{Affil:ioacam}}{Institute of Astronomy, University of Cambridge, Madingley Road, Cambridge, CB3 0HA, UK}
\altaffiltext{\ref{Affil:kavmit}}{Department of Physics, Kavli Institute for Astrophysics and Space Research, Massachusetts Institute of Technology, Cambridge, MA 02139, USA}
\begin{abstract}
We confirm that the object DDO216-A1 is a substantial globular cluster at the center of Local Group galaxy DDO216 (the Pegasus dwarf irregular), using Hubble Space Telescope ACS imaging. By fitting isochrones, we find the cluster metallicity [M/H] = $-$1.6 $\pm$0.2, for reddening E(B$-$V) = 0.16$\pm$0.02; the best-fit age is 12.3$\pm$0.8~Gyr.  There are $\approx$30 RR~Lyrae variables in the cluster;  the magnitude of the fundamental mode pulsators gives a distance modulus of 24.77 $\pm$0.08 --- identical to the host galaxy.  The ratio of overtone to fundamental mode variables and their mean periods make DDO216-A1 an Oosterhoff Type I cluster. We find a central surface brightness 20.85$\pm$0.17 F814W~mag~arcsec$^{-2}$, a half-light radius of 3$\farcs$1  (13.4~pc), and an absolute magnitude M$_{814}$ = $-$7.90 $\pm$0.16 (M/M$_{\sun}$ $\approx$10$^5$). King models fit to the cluster give the core radius and concentration index, r$_c$ = 2$\farcs$1$\pm$0$\farcs$9 and $c$ = 1.24$\pm$0.39. The cluster is an ``extended'' cluster somewhat typical of some dwarf galaxies and the outer halo of the Milky Way. The cluster is projected $\lesssim$30~pc south of the center of DDO216, unusually central compared to most dwarf galaxy globular clusters. Analytical models of dynamical friction and tidal destruction suggest that it probably formed at a larger distance, up to $\sim$1~kpc, and migrated inward. DDO216 has an unexceptional cluster specific frequency, S$_N$ = 10. DDO216 is the lowest-luminosity Local Group galaxy to host a 10$^5$~M$_{\odot}$ globular cluster, and the only transition-type (dSph/dIrr) in the Local Group with a globular.
\end{abstract}
\section{Introduction}
\setcounter{footnote}{0}

Dwarf galaxies (M$_* \lesssim 10^8 M_{\sun}$) are the most abundant class of galaxies in the Universe. They occupy an important part of parameter space for understanding the feedback processes that seem to govern the relationships between dark halo mass, baryon fraction, and star formation efficiency. Furthermore, their progenitors at high redshift may have played an important role in reionizing the Universe \citep{rob13}. Despite the ubiquity of dwarf galaxies, it is challenging to reliably measure their physical properties and put them in their appropriate cosmological context \citep[e.g.,][]{boy15}. 

Most dwarf galaxies are undetectable beyond redshift $z$ $\approx$1--2, even in the Hubble Ultra Deep Field or with planned JWST observations \citep{boy16}. Thus, observations of Local Group galaxies have set the benchmark for the accuracy and precision with which ancient star formation rates and chemical evolution histories can be measured
\citep[e.g.,][and references therein]{ski14,col14,wei14}. Long-lived main-sequence stars born at lookback times $\geq$10~Gyr, corresponding to $z$ $\gtrsim$ 2, are a direct probe of galaxy evolution in the Universe from the earliest star-forming period, through the epoch of reionization and its aftermath. 

In this paper we present a photometric analysis of the understudied star cluster at the center of DDO~216 (the Pegasus dwarf irregular, PegDIG, UGC~12613), which we observed serendipitously during our HST program to measure the complete star formation history of this galaxy. PegDIG (M$_V$ = $-$12.5 $\pm$0.2) is roughly a magnitude fainter than the Fornax and Sagittarius dwarfs, which makes it one of the least luminous galaxies known to host a cluster near the peak of the globular cluster luminosity function \citep[See][for examples of similar clusters in dwarfs with M$_V$ $\approx -$11.5]{geo09a,dac09}.

First, we review the basic parameters of the galaxy, and then describe our observations and reductions in section~\ref{sec:obs}. Our analysis of the structure and content of the star cluster DDO216-A1, including age and metallicity estimates based on the color-magnitude diagram and an analysis of the variable star population of the cluster, is given in section~\ref{sec:clus}.  We place DDO216-A1 in context with the population of massive star clusters in dwarf galaxies and summarise our results in section~\ref{sec:summ}. 

\subsection{Globular Clusters in Dwarf Galaxies}

\begin{figure*}[t!]
\plotone{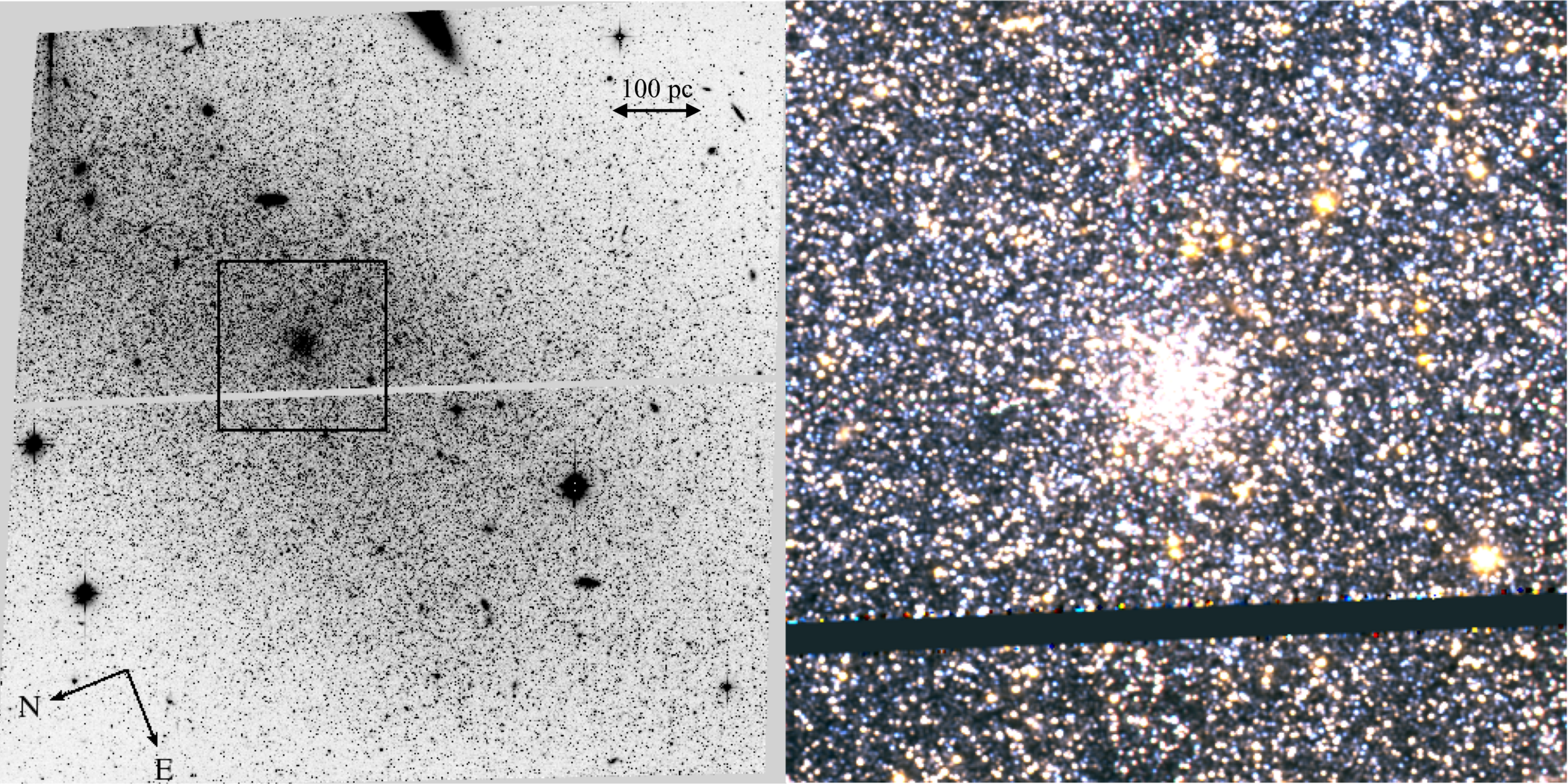}
\caption{DDO216 and its globular cluster. {\it Left:} Coadded ACS/WFC image of DDO216 through filter F814W. The total exposure time is 34.3 ksec, spread across 4 days. The cluster DDO216-A1 is clearly visible, 6$\arcsec$ south of the galaxy center. The box around the cluster is 45$\arcsec$ ($\approx$200~pc) across. {\it Right:} Magnified view of the 45$\arcsec$ region around DDO216-A1, constructed from F475W and F814W images. The cluster is high-surface brightness, dominated by red giants and horizontal branch stars, and by eye appears to be $\sim$8-10$\arcsec$ ($\sim$35--44~pc) in diameter.
\label{fig-image}}
\end{figure*}

Given deep enough observations, it is relatively straightforward to derive star formation rates at high lookback times for galaxies within the Local Group. It is more difficult to identify the triggers of star formation. For example, the majority of dwarf-dwarf major mergers are expected to have occured in the first few billion years after galaxy formation began \citep*[e.g.,][]{dea14}, but because of the destructive nature of mergers most of the obvious evidence for merger activity will have long since vanished. 

Globular clusters are an important window into this process because they require extreme conditions to form, suggestive of vigorous star formation in the mode often associated with galaxy mergers and interactions \citep[e.g.,][and references therein]{bro06}. Globular clusters are tightly bound and will typically survive for a Hubble time unless disrupted in a hostile tidal environment, but the relatively shallow potential wells of dwarf galaxies are not generally conducive to cluster disruption.


As a result, many globular and open clusters are known in dwarf galaxies in the Local Group and beyond, in enough numbers to make statistical associations between properties like host galaxy morphology and cluster colours and sizes \citep*[e.g.,][]{sha05,mil07}. These span a range of sizes from extremely luminous and dense nuclear star clusters to low-mass open cluster or association analogues, in galaxies down to some of the least luminous known. 

In the Local Group, recent work has discovered examples of modest star clusters even among the smallest galaxies \citep[e.g.,][]{crn16}, and more massive, sometimes extended clusters in some of the larger irregular galaxies \citep{sha07,hwa11}. In light of these discoveries and others, \citet*{zar16} have suggested that some of the outer halo globular clusters thought to be Galactic globular clusters may in fact be hosted by undiscovered low surface brightness galaxies. However, the lowest-luminosity Local Group galaxies with catalogued globular clusters similar to the massive and dense globulars of the Milky Way are Fornax and Sagittarius (M$_V$ = $-$13.4 and $\lesssim -$13.5, respectively). 

\subsection{The Pegasus Dwarf Irregular, DDO 216}
\label{sec:peg}

The Pegasus dwarf irregular galaxy, PegDIG, was discovered by A.G. Wilson in the early 1950's on Palomar Schmidt plates \citep{hol58}. From early on it was considered to be a candidate member of the Local Group with a distance of $\sim$1~Mpc. The case for membership was supported by the negative heliocentric \ion{H}{1} radial velocity found by \citet{fis75}. PegDIG is considered a distant M31 satellite \citep[d$_{\mathrm M31}$ $\approx$470~kpc,][]{mcc07}; it is not proven that it has ever interacted with M31, although it is more likely than not that PegDIG has previously been within M31's virial radius \citep{sha13,gar14}.
PegDIG is fairly isolated at the present time, its nearest neighbor being the M31 satellite And~VI, just over 200~kpc away. It is quite unlikely that PegDIG has had strong tidal interactions with any other known galaxy during the past several Gyr.

PegDIG is a fairly typical small irregular galaxy, with roughly 1:1 gas to stellar mass ratio, although it has virtually no current star formation as measured by H$\alpha$ emission \citep{you03}.  This leads to its classification as a transition type dwarf, with properties intermediate between the spheroidal (dSph) and irregular (dIrr) types \citep{mcc12}. It has an ordinary metallicity of [Fe/H] $\approx$ $-$1.4 $\pm$0.3 for its stellar mass of $\approx$10$^{7}$~M$_{\odot}$ \citep{kir13}. Unlike the spheroidal galaxies, it appears to be rotating; both HI \citep{kni09} and stellar \citep{kir14} data suggest a rotation speed (not corrected for inclination) of $\approx$15--20~km/s. \citet{mcc07} drew attention to the cometary appearance of the neutral gas, and attributed the asymmetric morphology to ram pressure stripping by diffuse gas in the Local Group, although this conclusion is disputed, based on much deeper HI observations, by \citet{kni09}. 

The star formation history of PegDIG has been estimated from ground-based data reaching a limiting absolute magnitude of M$_{\mathrm I}$ $\approx$ $-$2.5 \citep*{apa97}, and from HST/WFPC2 observations \citep{gal98} reaching $\approx$2.5 mag deeper.
Within large uncertainties \citep{wei14a}, the picture that emerges from these studies is of star formation that has spanned a Hubble time, likely to be declining over time following an early epoch of high star formation rate (SFR). The SFR has certainly declined with time over the past $\approx$1--2~Gyr, despite the large reservoir of neutral gas.

In its extended star formation history, PegDIG appears to have more in common with the dIrr galaxies than with a typical dSph, consistent with its retention of neutral gas to the present day and with the assertion of \citet*{ski03} that transition type galaxies represent the low-mass/low-SFR end of the dwarf irregular population \citep[see also][]{wei11}. The precise star formation history over the full lifetime of the galaxy will be determined in a future paper in this series (Cole et al., in preparation).

\section{Observations \& Data Reduction} \label{sec:obs}

We observed PegDIG using the Advanced Camera for Surveys Wide Field Camera (ACS/WFC) as part of the Cycle 22 program GO-13768.  The observations, which comprise 34.3 and 37.4 kiloseconds in the F814W (Broad I band) and F475W (Sloan $g$ band) filters respectively, were made between 23--26 July, 2015. 29 orbits were allocated to the Pegasus observations, split into 15 visits of one to two orbits each to facilitate the detection of short-period variable stars. Each orbit was broken into one exposure in each filter. Parallel observations at a distance of $\approx$6$\arcmin$ were obtained simultaneously through the equivalent filters on the Wide Field Camera 3.
 
The charge-transfer efficiency corrected images were processed through the standard HST pipeline and photometry was done using the most recent version of DOLPHOT, with its HST/ACS-specific modules \citep{dol00}. Extended objects and residual hot pixels were rejected based on their brightness profiles, and aperture corrections were derived based on relatively isolated stars picked from around the image. Stars that were found to suffer from excessive crowding noise (crowding parameter $>$1.0) due to partially resolved bright neighbors were rejected, leaving a sample of 247,390 well-measured stars with S/N $\geq$5.  We performed an artificial star test analysis with 50,000 stars, distributed following the cluster light within 10$\arcsec$ of the cluster center, in order to characterize the measurement errors and incompleteness. The 50\% completeness limits within the approximate cluster radius of 5$\arcsec$ are (m$_{475}$, m$_{814}$) = (27.1, 25.7);
this does not reach the cluster MSTO due to crowding, although the oldest MSTO is reached for the field population.  
At these magnitude levels the typical photometric error is $\lesssim$0.03~mag, allowing us to resolve the old stellar sequences and make comparisons to isochrone models with a high degree of precision. 

The co-added image built from our F814W images is shown in Figure~\ref{fig-image}a. This tends to highlight the foreground stars and many background galaxies, along with the smoothly varying light of the intermediate-age and old  field population of PegDIG. There is little to no evidence for associations of luminous, young stars, consistent with the evidence for a very low rate of massive star formation \citep*{ski97}. The distribution of bright stars is clumpy and irregular, with a few modest dust lanes that could contribute to differential reddening.

DDO216-A1 is highlighted within a 45-arcsecond box in Figure~\ref{fig-image}a. Fig.~\ref{fig-image}b expands this highlighted region in false color. The gap between WFC1 and WFC2 chips runs across the bottom third of the cluster image. The cluster is prominent in the image, even seen in projection against the densest part of the galaxy field population. The field population is patchy and irregular, with evidence for differential reddening and an increase in the density of bright, blue stars towards the upper right.  

The cluster appears to be around 8--10~arcseconds (35--44~pc) in diameter, and is nearly circular in projection. DDO216-A1 is neither notably blue nor red compared to its surroundings. The neutral color indicates immediately that the cluster light is dominated by first-ascent red giant branch stars, and not young main-sequence stars or asymptotic giant branch stars. The cluster structural and photometric properties are presented in Section~\ref{sec:clusphot}.

\begin{figure}[t!]
\plotone{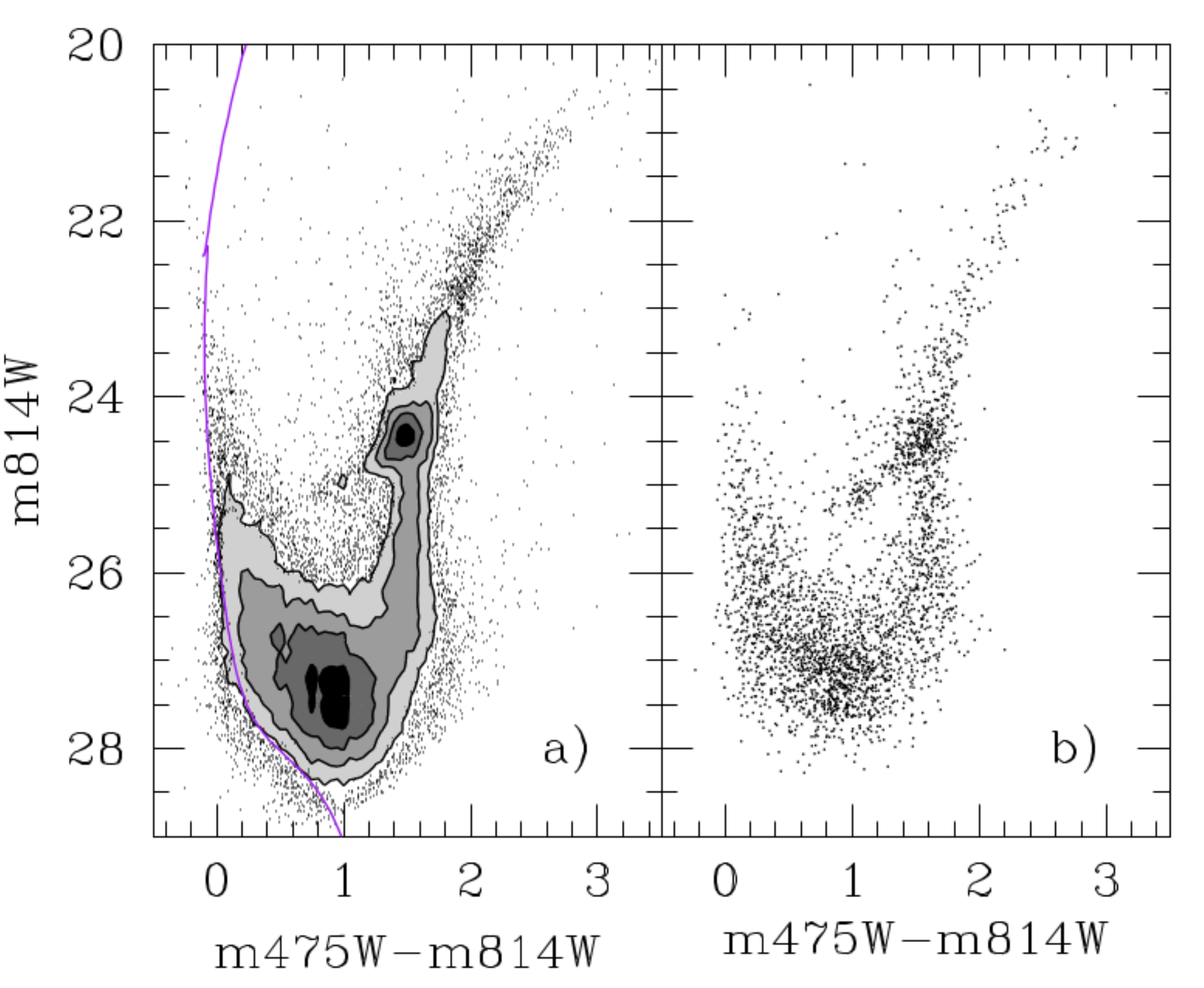}
\caption{ACS/WFC color-magnitude diagrams of the central part of DDO~216. {\it a)} The central arcminute of the galaxy, with the CMD binned 0.05$\times$0.1mag and contoured at 30, 60, 120, 240 stars per bin; the galaxy contains stars spanning a wide range of ages. A 30~Myr PARSEC isochrone for metallicity Z = 0.001 has been overlaid for m$-$M$_0$ = 24.77, E(B$-$V) = 0.16; {\it b)} The central 10$\arcsec$ around DDO216-A1. Incompleteness is much higher due to increased crowding; numerous field stars are present but the cluster CMD is dominated by a horizontal branch at (m475W$-$m814W) $\approx$ 1 and a well-populated red giant branch, lacking main-sequence stars above m814W $\lesssim$27.5 and red clump stars relative to the field.
\label{fig-cmd}}
\end{figure}

We present a CMD of the central 1 arcminute of the galaxy (including the cluster) in Figure~\ref{fig-cmd} to highlight the differences between the galaxy field and the cluster. Compared to its surroundings, the cluster suffers from a much higher degree of crowding, but is still complete to at least a magnitude below the level of the horizontal branch. A 30~Myr PARSEC isochrone \citep{bre12} chosen to approximately match the mean metallicity of PegDIG (Z = 0.001) has been overlaid on the field CMD to indicate the distance, 900 $\pm$30~kpc, and reddening, E(B$-$V) = 0.16 $\pm$0.02~mag to the galaxy (see below).

\begin{figure}[t!]
\plotone{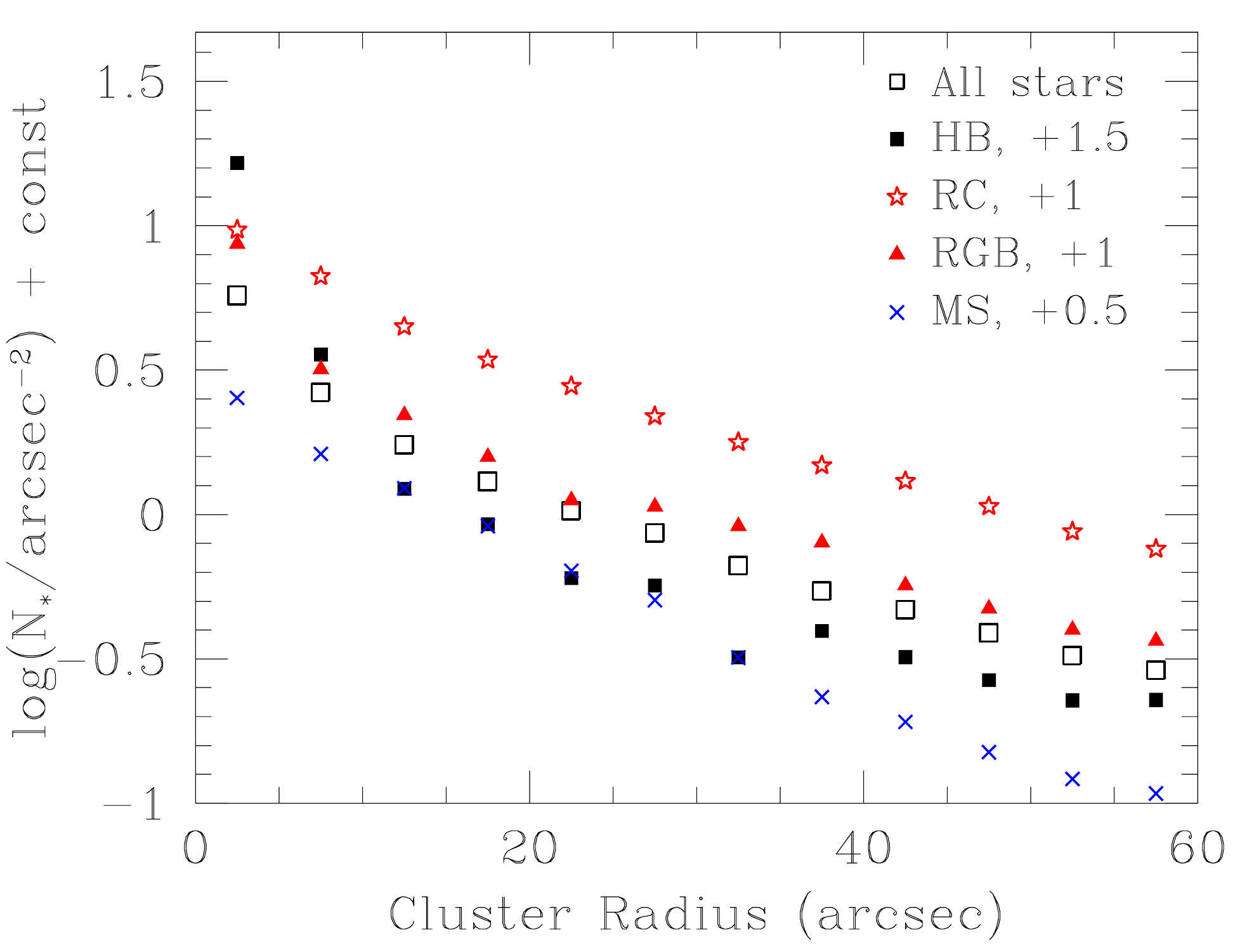}
\caption{Radial density profile of stellar populations over 1~arcminute surrounding the cluster. An arbitrary offset has been applied to the logarithm of the surface densities to facilitate comparison, listed in the figure legend. All stars brighter than m814 = 26 are plotted as open squares. The horizontal branch (HB) shows the strongest increase in the area around the cluster, followed by the red giant branch (RGB). The red clump (RC) profile has a similar slope to the HB and RGB in the outer regions, but steepens less in the center, while the main sequence (MS) shows no evidence of central steepening. See text for the color cuts that define the selection regions.
\label{fig:propops}}
\end{figure}

The field stars contain a strong population of red clump stars, a relatively broad red giant branch, and main sequence that is well-populated across a wide range of magnitudes, indicating a wide range of stellar ages. In order to test whether the cluster represents a distinct stellar population from the field or simply a high-density core drawn from the surroundings, we plot the radial density distribution of stars in different parts of the CMD in Figure~\ref{fig:propops}. Only bright stars have been considered in order to avoid problems with radial variation of crowding. The overall profile is nearly consistent with a power law over the entire inner arcminute of the galaxy, with an upturn in the inner 10$\arcsec$. Various sub-populations have been overplotted, with offsets applied to facilitate comparison of the radial profiles. The density of horizontal branch (HB) stars with 0.8 $\leq$ (m475$-$m814) $\leq$ 1.2 rises very steeply within 10$\arcsec$, while the red clump (RC) stars, with 1.3 $\leq$ (m475$-$m814) $\leq$ 1.8 show a much less pronounced increase. The red giant branch (RGB) profile for stars brighter than m814 = 24 has a similar slope to the HB and RC at large radii, but steepens at a rate between the HB and the RC in the cluster center. The upper main sequence profile, for stars with (m475$-$m814) $\leq$ 0.4, is steeper than the bulk profile overall, but does not further steepen in the center.

The cluster shows only slightly increased density of red clump stars, but shows a well-populated HB and RGB, with a mostly red horizontal branch morphology. The tip of the red giant branch in the cluster terminates at m814W $\approx$21, very similar to that of the field. The main-sequence turnoff (MSTO) of the cluster is not obvious due to the increased crowding; if the cluster were younger than $\approx$7--8~Gyr we would see its MSTO above the crowding limit of the data.  Note, however, that Fig.~\ref{fig:propops} only constrains the MS density for stars younger than $\approx$3--4~Gyr. We further quantify the cluster age and metallicity in Section~\ref{sec:agemet}, below.

\section{Globular Cluster DDO216-A1} \label{sec:clus}

\subsection{Historical Observations}

The first CCD images of PegDIG were obtained by \citet{hoe82}, who noted three star cluster candidates and also commented on the high number of background galaxies visible around and through the galaxy. Indeed, PegDIG is not far from the supergalactic plane; it appears in projection on the sky in the foreground of a galaxy group in the outskirts of the Pegasus cluster \citep[see e.g.,][]{chi76,gal98,kri01}. \citet{hoe82} listed their star cluster candidates as A1--3; our ACS/WFC imaging shows candidates A2 and A3 to be background galaxies, but we confirm A1 to be a bona fide cluster. The potential for naming confusion with the Pegasus cluster of galaxies leads us to adopt Hoessel \& Mould's nomenclature for the central star cluster in DDO216, which we hereafter refer to as DDO216-A1. \citet{hoe82} report the magnitude and color for cluster A1, in the Thuan-Gunn system, as $G$ = 18.75 and $G-R$ = 0.45, and estimate its age to be $\gtrsim$~2~Gyr from the integrated color.

The first Hubble Space Telescope observations of PegDIG were reported by \citet{gal98}, who placed the WFPC2 camera in the central parts of the galaxy such that the PC chip barely included the outskirts of DDO216-A1. The cluster is also clearly identifiable in their 0\farcs6-seeing WIYN~3.5m telelecope images.  These authors note that the cluster is only moderately dense, with a rather large diameter of $\approx$40~pc, and that although it is located very centrally within PegDIG it falls well short of their expectations for a galaxy nucleus or super star cluster \citep*{oco94}. The shallowness of their CMDs and the positioning of the cluster mainly outside the WFPC2 field prevented any further quantitative work although they confirm the claim of \citet{hoe82} that the cluster is most likely older than 2~Gyr, based on its lack of bright main-sequence stars \citep{gal98}.

Subsequently the cluster appears to have been lost to the literature, failing to be included in compilations of the properties of Local Group galaxies
\citep[e.g.][]{mat98,for00},
papers specifically devoted to the statistics of star clusters in dwarf galaxies \citep*[e.g.,][]{bil02,sha05,geo09a}, and further space- and ground-based studies of PegDIG \citep[e.g.,][]{tik06,mcc07,boy09}. Given the recent surge in interest in the nature of extended star clusters in dwarf galaxies and in the physical differences between clusters and galaxies, the presence of an understudied, luminous, high surface brightness star cluster at the very center of one of the Local Group's faintest photographically discovered galaxies is remarkable.

\subsection{Cluster Size and Luminosity} \label{sec:clusphot}

\begin{figure}[t!]
\plotone{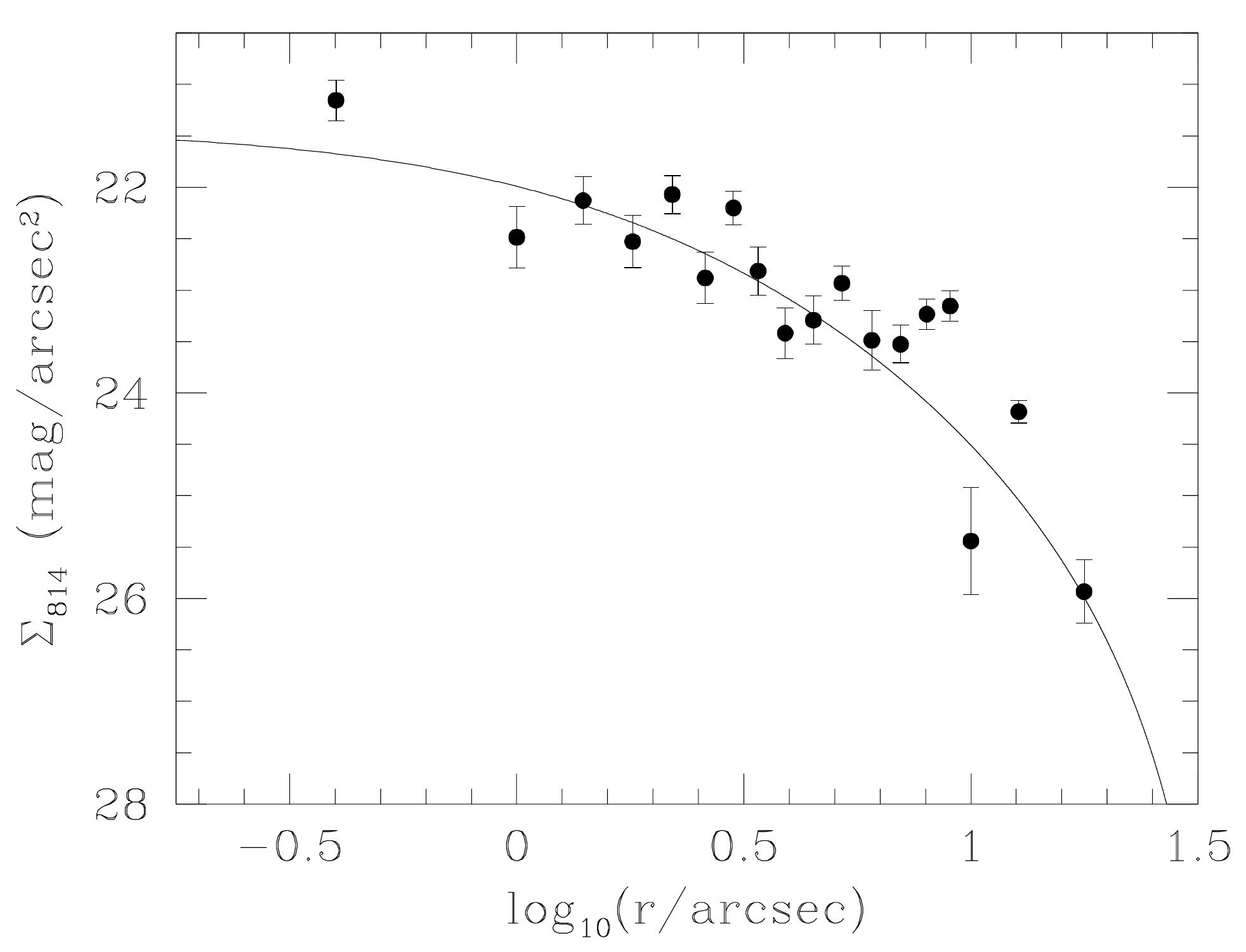}
\caption{F814W radial surface brightness profile of DDO216-A1, based on aperture photometry. A fit to a
King model profile with central surface brightness $\Sigma_{814}$ = 20.85 mag arcsec$^{-2}$, core radius
r$_c$ = 2.12$\arcsec$, and concentration index $c = 1.24$ is shown.
\label{fig-profile}}
\end{figure}

DDO216-A1 is located at right ascension 23:28:26.3, declination $+$14:44:25.2 (J2000.0).  This is just 6$\arcsec$ (26 pc) south of the center of the outer red light isophotes of PegDIG as measured by \citet{mcc07}. The cluster is of high surface brightness and has a very distinctly identifiable core, but determination of the cluster properties is complicated by the fact that it is embedded in the highest surface brightness part of PegDIG, and by the fact that significant differential reddening appears to be present. The differential reddening can be seen both in the color spread of red giants and upper main sequence stars, and in the presence of dust lanes in the false color ACS/WFC images. DDO216-A1 is nearly circular; using the IRAF\footnote{IRAF is distributed by the National Optical Astronomy Observatories, which are operated by the Association of Universities for Research in Astronomy, Inc., under cooperative agreement with the National Science Foundation.} {\it stsdas} task {\it ellipse}, we find an ellipticity $\epsilon$ $\equiv$ (1$- b/a$) $<$0.15. 

We calculated the half-light radius of the cluster using star counts of horizontal branch stars. The PegDIG field has a compound HB morphology, with both red and blue HB stars present, spread nearly evenly across the ACS/WFC field. The cluster core, within 40 pixels of the center, has roughly an order of magnitude higher surface density of HB stars, so these make a convenient measure for estimating the cluster extent with the minimum field contamination.

In a region from 12--15$\arcsec$ outside the cluster, the density of HB stars is 0.14 arcsec$^{-2}$. The HB density rises to twice this value at a radius of 8$\farcs$5 from the cluster center, which we take to be the area over which the cluster is clearly dominant over the field. Within this 8$\farcs$5 ($\approx$37~pc) circle, we would expect to find 32 $\pm$6 HB stars from the field population alone; the actual CMD of this area yields 89 HB stars. Subtracting the field contribution and finding the radius within which half the cluster is contained, a half-light radius of 3$\farcs$1 $\pm$ 0$\farcs3$ is derived (13.4 $\pm$ 1.3~pc). This result does not significantly change if the upper red giant branch stars are included, and the contamination by field stars is higher. 

We measured the flux in concentric apertures around the cluster, using annuli with outer radii from 0$\farcs$8--20$\arcsec$ in size, and an annulus from 25--30$\arcsec$ to measure the field contribution. We measured the F814W and F475W surface brightness profiles and found them to be consistent with the half-light radius derived from HB star counts. 

We merged the two bandpasses into a single profile by simultaneously fitting a King model with a common core radius r$_c$ and concentration index $c$ $\equiv$ log(r$_t$/r$_c$) and two central surface brightness values, one for the F814W image and one for the F475W image. The resulting fit is shown for the F814W surface brightness in Figure~\ref{fig-profile}. The King model core radius is r$_c$ = 2$\farcs$12 $\pm$ 0$\farcs$91, the concentration index is $c$ = 1.24 $\pm$0.39, and the central surface brightnesses are $\Sigma_{814}$ = 20.85 $\pm$0.23 and $\Sigma_{475}$ = 22.68 $\pm$0.24 mag~arcsec$^{-2}$, respectively. 

The outer regions of the cluster are difficult to reliably measure because of the high field contamination. Therefore, the total magnitude of DDO216-A1 is estimated by measuring the background-subtracted flux within the half-light radius and doubling it. The extrapolated integrated magnitudes are (m475W, m814W) = (18.64 $\pm$0.07, 17.17 $\pm$0.14). Transformation to a standard Johnson-Cousins system gives (B, I) = (18.95, 17.39), and a transformed V magnitude of 18.13. The measured total magnitude is thus in reasonable accord with the original discovery values in \citet{hoe82}. 

The absolute magnitude of DDO216-A1 is M$_{814}$ = $-$7.90 $\pm$0.16 for the distance and reddening derived from our CMDs of the cluster and its surrounding fields. This yields a V magnitude M$_{\mathrm V}$ = $-$7.14, slightly fainter than the peak of the Milky Way globular cluster luminosity function. The color B$-$V = 0.82 is quite red for a globular cluster, but when dereddened [(B$-$V)$_0$ = 0.66] is entirely consistent with typical metal-poor globulars.

\begin{figure}[t!]
\plotone{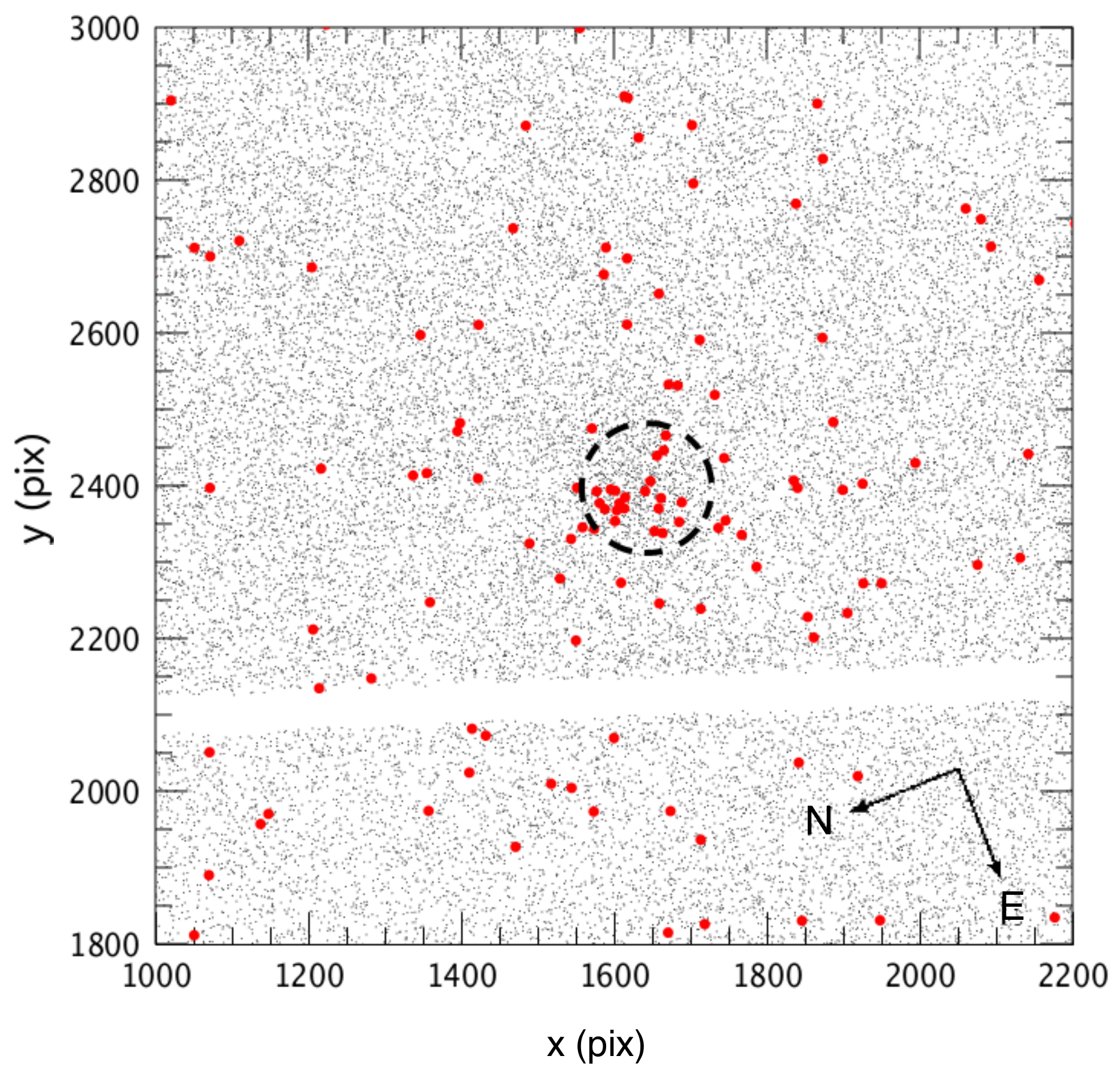}
\caption{Distribution of RR~Lyrae variables in the central 1~arcmin$^2$ of PegDIG. Non-variable stars are plotted in grey; the variables are heavy, red circles. The dashed circle centered on the cluster DDO216-A1 has a diameter $d$ = 8$\arcsec$.
\label{fig:rrmap}}
\end{figure}

\begin{figure}[h]
\plotone{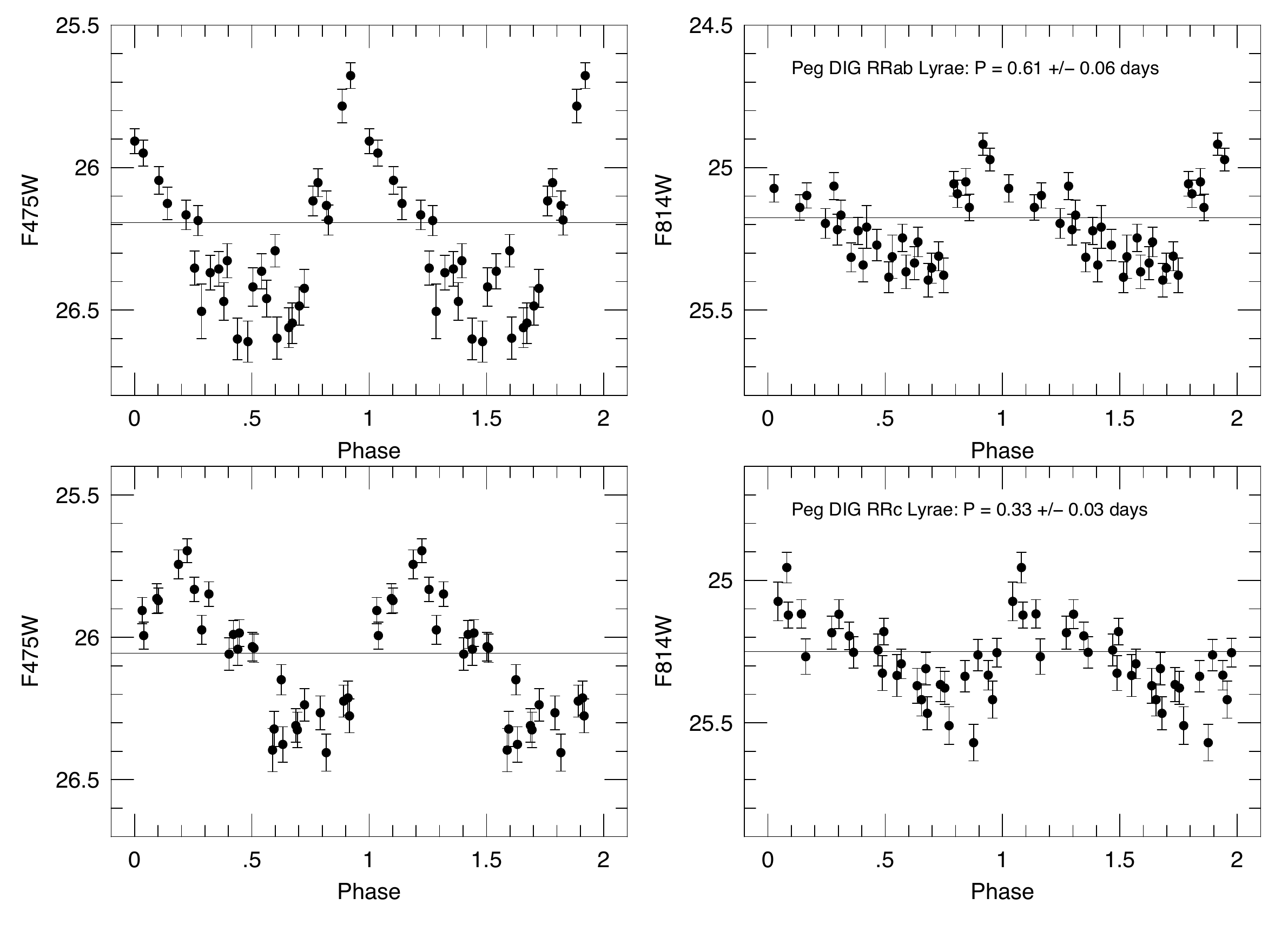}
\caption{Sample RR~Lyrae type variable star light curves for DDO216-A1. {\it Top row:} star 39890, type RRab; {\it left:} F475W; {\it right:} F814W. {\it Bottom row:} As above, for star 39268, type RRc.
\label{fig:rrlyr}}
\end{figure}

\subsection{Variable Stars and Distance}

Candidate variable stars were identified in the photometric catalog using the method of \citet{sah90}, as implemented and applied to {\it HST} photometry in \citet{dol01} and numerous subsequent papers \citep[e.g.,][and references therein]{mcq15}. The algorithm labels stars as likely variables if the photometric scatter of the individual data points is much larger than the photometric errors, the scatter is not due to a small number of outliers, and the light curve is periodic, with a best-fit period found through an application of the \citet{laf65} algorithm.  The observations are highly sensitive to variables of period from $\approx$0.1--3 days, with 29 individual data points in each filter, spread over 2.8 days with an average gap between observations of 0.05 days.

Over the entire ACS/WFC field, several hundred likely variables were flagged, strongly clustered on the horizontal branch in the CMD, with periods of $\approx$0.5--0.6~day. A scattering of brighter, longer period variables, mostly within the Cepheid instability strip, was also identified.  Figure \ref{fig:rrmap} shows the location of all stars in the central 1~arcmin$^2$ portion of the galaxy.  RR~Lyrae variables are marked with larger, red points; the dashed circle at the position of the cluster has a diameter of 8$\arcsec$. The horizontal branch variables are found to be distributed evenly across the galaxy, with the exception of a strong concentration at the location of DDO216-A1. The field variable star population of PegDIG will be analysed in a future paper (Skillman et al., in preparation). 

Here, we identify genuine variables within 8 arcseconds of the cluster center and classify them using their light curve shapes.
Within this area, there are 32 variables, all on or near the horizontal branch; the corresponding number in a typical nearby comparison region is 4, so the vast majority of the 32 are likely to be cluster members.  Among the cluster variables, 27 are fundamental mode (RRab) pulsators and 5 are overtone (RRc) pulsators. Sample light curves in each filter for one star of each variable type are shown in Figure~\ref{fig:rrlyr}, with the period and mean magnitude shown. 

Four additional variables with brighter magnitude but unexceptional periods and colors were also found, $\approx$0.1--0.4~mag above the HB. Their colors are too blue to be population II Cepheids; if they were the descendents of a putative blue straggler population, they would also likely fall to the red of their observed position in the CMD. These are likely to be photometric blends; for now, we exclude them from further analysis of the cluster properties. The 50\% completeness limit is $\approx$1~magnitude below the HB in the cluster region, so the census of cluster variables is likely to be close to complete (apart from the 4 rejected blends). The list of variables within 8$\arcsec$ of the cluster center is given in Table~\ref{tab:rrlyr}.

\begin{deluxetable*}{lcccccr}
\centering
\tablecolumns{7}
\tablewidth{0pt} 
\tablecaption{Variable Stars in DDO216-A1\tablenotemark{$\dagger$} \label{tab:rrlyr}}
\tablehead{
\colhead{ID} & 
\colhead{$x$ (pix)} &
\colhead{$y$ (pix)} &
\colhead{$\langle$m475W$\rangle$} &
\colhead{$\langle$m814W$\rangle$} &
\colhead{Period (d)} &
\colhead{Type}}
\startdata
39184 & 1543.92 & 2330.24 & 26.103$\pm$0.074 & 25.141$\pm$0.036 & 0.500$\pm$0.065 & RRab \\
39321 & 1551.63 & 2397.23 & 26.156$\pm$0.056 & 25.176$\pm$0.033 & 0.559$\pm$0.075 & RRab \\
38807 & 1558.91 & 2345.44 & 26.152$\pm$0.055 & 25.134$\pm$0.034 & 0.559$\pm$0.072 & RRab \\
40423 & 1571.24 & 2475.00 & 26.284$\pm$0.069 & 25.208$\pm$0.034 & 0.563$\pm$0.068 & RRab 
\enddata
\tablenotetext{$\dagger$}{A sample of the table is provided here for guidance regarding the form and content. The complete table may be found in the online version of the paper.}
\end{deluxetable*}

\begin{figure*}[t!]
\plotone{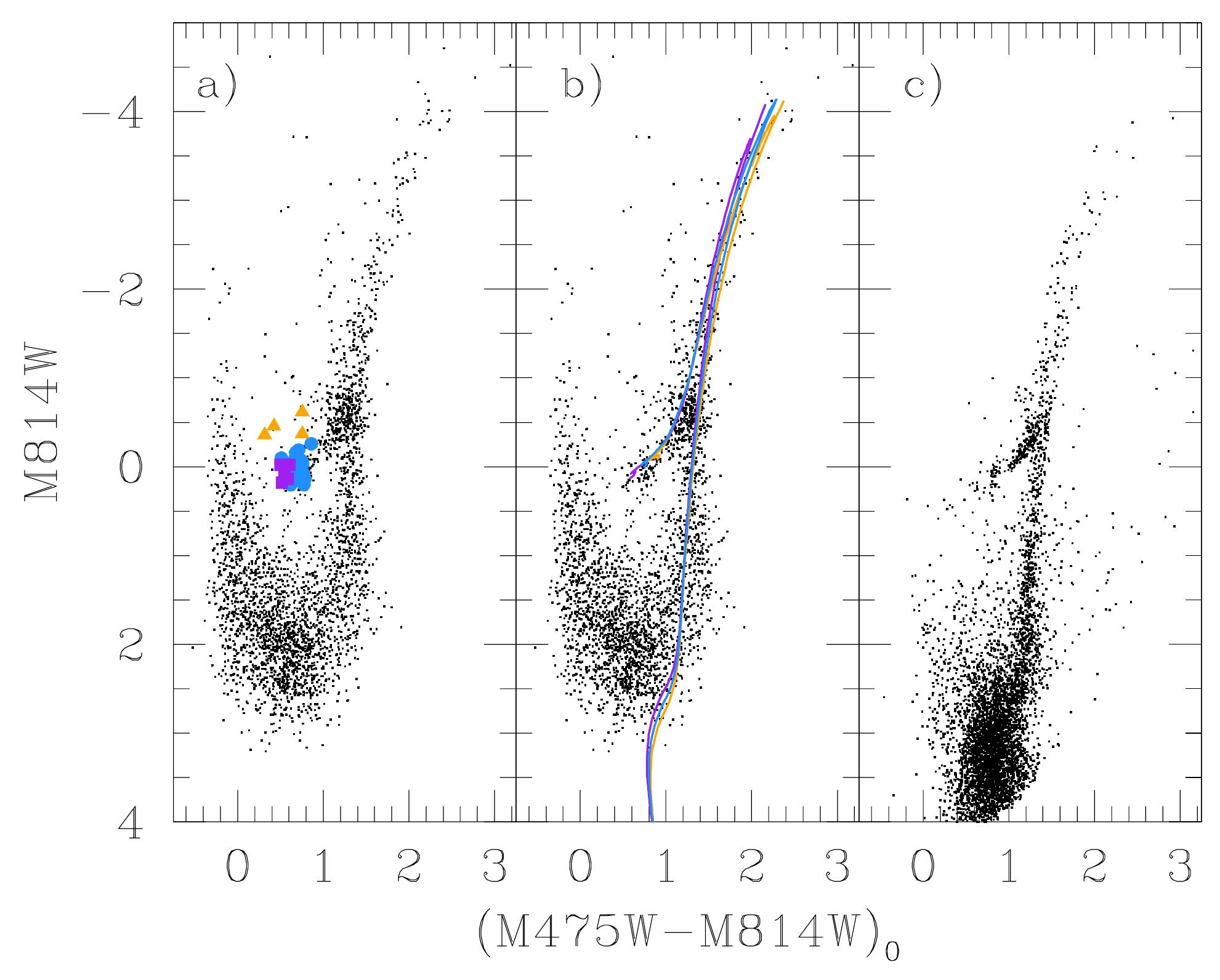}
\caption{Color-magnitude diagrams of the cluster (left and center panel) and a comparison field far from the center of PegDIG. {\it a)} CMD of stars within 8$\arcsec$ of DDO216-A1, with all variable stars highlighted. Those classified as RRab are shown as blue circles, RRc as purple squares, and unclassified variables as orange triangles. The photometry has been corrected for (m$-$M)$_0$ = 24.77, E(B$-$V) = 0.16; {\it b)} Photometry as for panel {\it a}, but with three PARSEC isochrones overlaid: (Z/10$^{-4}$, age/Gyr, color) = (2.4, 11.5, purple), (4, 12.3, blue), (6, 13.2, orange); {\it c)} Comparison field showing the stars more than 1$\farcm$8 ($\approx$500~pc) south of the cluster. The lack of both crowding and differential reddening result in greater photometric depth and tighter stellar sequences. The photometry in this panel has been corrected for a smaller reddening, E(B$-$V) = 0.07, consistent with pure foreground dust. 
\label{fig-cmdplus}}
\end{figure*}

The variable stars of DDO216-A1 are overplotted on the cluster CMD in Figure~\ref{fig-cmdplus}a. The magnitudes have been corrected for a reddening E(B$-$V) = 0.16 as derived from the color of the upper main sequence, and for distance modulus of 24.77, derived from the mean magnitude of the fundamental mode pulsators and isochrones fits to the cluster (see below). 

The mean period of the 27 RRab pulsators that are likely members of DDO216-A1 is 0.556 $\pm$0.039~day, and
the mean period of the 5 probable RRc members is 0.356 $\pm$0.012~day.  This, along with the roughly 5:1 ratio of fundamental to overtone pulsators, identifies the cluster as belonging to Oosterhoff Type I \citep[Oo~I,][]{oo39}.
Most Galactic globular clusters more metal-rich than [Fe/H] $\approx$ $-$1.7 fall into this category, with 
$\langle$P$\rangle_{ab}$ $\approx$0.55~d; most of the more metal-poor clusters are type Oo~II, with 
$\langle$P$\rangle_{ab}$ $\approx$0.65~d \citep{cat15}. Interestingly, many of the globular clusters in the Fornax and Sagittarius dSph and in the Large Magellanic Cloud are of intermediate or indeterminate Oosterhoff type \citep[e.g.,][]{mac03,sol10}; whatever effect causes these differences in variable populations seems not to occur for DDO216-A1. DDO216-A1 appears to be quite ordinary in its variable star population; normalised to a cluster with M$_V$ = $-$7.5, the specific frequency of RR~Lyraes is $\approx$42.

The mean F475W and F814W magnitudes of the probable RRab cluster members are 26.074 $\pm$0.126 and 25.088 $\pm$0.111, respectively. We convert the magnitudes to a mean V magnitude following the procedure in \citet{ber09}, finding $\langle$V$\rangle$ = 25.699; the rms scatter is $\pm$0.114 and the standard error of the mean is $\pm$0.022. There is significant scatter in the light curves, but the mean amplitude (F475W) of the RRab variables is $\approx$0.9~mag, consistent with the identification as a type Oo~I cluster.

The mean RRab magnitude can be used to derive the distance, provided some estimate of the metallicity is known \citep{san90,dem00}.  One choice is to adopt the mean metallicity and metallicity spread from the spectroscopic study of PegDIG red giants by \citet{kir13}, [Fe/H] = $-$1.4 $\pm$0.3. Isochrone fits to the cluster red giant branch using the PARSEC isochrones \citep{bre12} give a slightly lower estimate, $-$1.6, with a random error $\pm$0.2 (see section~\ref{sec:agemet}). 

The metallicity estimates are mutually consistent within the errors, so we adopt the isochrone-based value since it is directly measured from the cluster and not from the field, $\gtrsim$100 parsecs away.  We use the metallicity-magnitude relationship with a zeropoint based on parallaxes of nearby field RR~Lyraes from \citet{ben11} to derive the variable star-based distance to DDO216-A1.  With [Fe/H]$_{\mathrm A1}$ = $-$1.6 $\pm$0.2 and M$_{\mathrm V}$(RR) = 0.214[Fe/H]$+$0.77, we find that the distance modulus to DDO216-A1, given the adopted reddening, is (m$-$M)$_0$ = 24.77$\pm$0.08. The uncertainty is dominated by the uncertainty in the V-band extinction. 

A complete analysis of the field star variable population is in progress, but here we note that the mean V-band magnitude of several hundred field RRab variables over the entire galaxy is $\langle$V$\rangle$ = 25.591 $\pm$0.093. If the reddening and the metallicity of the field population is taken to be identical to the cluster, then this would imply that the cluster lies some 43~kpc behind the rest of PegDIG. It is far more likely that the central region, including the cluster and the nearby field, is more highly reddened than the outer portions of the galaxy (Section \ref{sec:clusphot}). Indeed, the variables within the cluster footprint are redder than the average variable by $\Delta$(m475W-m814W) = 0.06~mag, lending support to this interpretation. The difference in reddening would cut any suggested distance offset more than in half. The field RR~Lyraes could also differ in metallicity from the cluster; if they were more metal-poor by 0.2--0.5~dex this would reconcile the mean V magnitudes, although this is not supported by metallicity estimates for PegDIG \citep{mcc05,kir13}.

The cluster distance, 899 $\pm$31~kpc, is identical to within the errors with the TRGB distance to the host galaxy, 919 $\pm$32~kpc, as determined by \citet{mcc05}, although it is formally slightly smaller. A direct comparison to the TRGB of PegDIG can be made using the current ACS/WFC data (paper in preparation). We find the TRGB in the central part of PegDIG to be m814W = 20.98 $\pm$0.11; we adopt the absolute magnitude of $-$4.06 from the calibration of \citet{riz07}. Combined with the reddening E(B$-$V) = 0.16 adopted here yields a TRGB distance modulus (m$-$M)$_0$ = 24.74 $\pm$0.15, virtually identical to the cluster RR Lyrae distance. If instead we use the lower reddening value from \citet{mcc05}, E(B$-$V) = 0.064, we would find a significantly higher distance modulus of 24.92.

In short, based on the redder colors and fainter magnitudes of its variable stars, it is likely that DDO216-A1 sits on the far side of PegDIG rather than the near side, but the preponderance of evidence does not indicate a significant distance offset.

\subsection{Age and Metallicity}
\label{sec:agemet}

We plot the CMD of the cluster core (r $\leq$ 8$\arcsec$) in Figure~\ref{fig-cmdplus}b. By fitting the distribution of stars using synthetic CMDs and artificial star tests we are able to constrain the age and metallicity of DDO216-A1. The CMD-fitting programs, {\it forge/anneal}, are described in \citet{ski03b,col07,col14}, and numerous other references. We adopt the most recent set of isochrones from the PARSEC family of models \citep{bre12}. These models have been calculated assuming the solar metallicity [M/H] = 0 corresponds to a mass fraction of elements heavier than helium Z = 0.01524. In the fits, we subtracted off a scaled Hess diagram drawn from an annulus around the cluster to account for field contamination.

The best-fit distance from the CMD is 24.80 $\pm$0.10, entirely consistent with the value derived from the variable stars. The reddening value is equal to the mean reddening of the upper main sequence stars in the nearby field, E(B$-$V) = 0.16 $\pm$0.02. We therefore choose to reduce the variance in our fits by holding these values fixed in our final determination of the cluster age and metallicity.

Because of the high degree of crowding, the cluster CMD becomes incomplete significantly brighter than the MSTO. There is thus some age/metallicity/reddening degeneracy in the best-fit solutions, as there always must be when the fit information is dominated by red giant branch stars. The best-fit isochrones have Z = 0.0004 $\pm$0.0002 ([M/H] = $-$1.6 $\pm$0.2), with age = 12.3 $\pm$0.8~Gyr (random error only). Covariance between age and metallicity mean that good models on the younger side of the range tend to have higher metallicity, and vice versa. Given the field contamination and crowding we are unable to rule out small numbers of younger stars or a cluster metallicity spread without spectroscopic information.

Figure~\ref{fig-cmdplus}b shows three isochrones overplotted from the range of acceptable solutions. The best-fit age and metallicity is plotted in blue, while the purple track shows Z = 0.0024 ([M/H] = $-$1.8), age = 11.5~Gyr, and the orange track gives Z = 0.0006 ([M/H] = $-$1.4), age = 13.2~Gyr. All three solutions clearly are reasonably good matches to the data, but the central part of the range gives the best reproduction of the cluster HB color.   

Detailed exploration of the field star CMD is beyond the scope this paper. However, we briefly consider the ancient field star populations as a point of comparison to the cluster. In Figure~\ref{fig-cmdplus}c we show the CMD of a region of the field far from the high surface brightness ``bar'' of PegDIG, in the southern corner of the ACS/WFC chip. These stars are all more than 1$\farcm$8 (500 pc) from DDO216-A1, a very low surface brightness part of the galaxy completely lacking in stars $\lesssim$2~Gyr old. This field, which has reddening consistent with E(B$-$V) = 0.07 \citep{sch11}, clearly shows a metal-poor, very old population similar to the globular cluster. The predominantly red HB morphology suggests a slightly more metal-rich population, consistent with the spectroscopic metallicity measurements from \citet{kir13}. 

Main sequence stars with M814W $\approx$ $+$1--2, and the hint of a ``vertical red clump'' \citep[VRC;][]{cap95} can be seen in Fig.~\ref{fig-cmdplus}c, indicating either a small population of intermediate-age, metal-poor stars, or ancient blue stragglers and their descendents.  The field blue stragglers and VRC give a clue about where to look in the cluster CMD for similar types of star; unfortunately both regions in the CMD of  DDO216-A1 are completely dominated by field populations of intermediate age.

The field VRC is at a color of $\approx$1.1 and a magnitude of $\approx$ $-$0.75, which is inconsistent with the location of the anomalously bright variable stars in DDO216-A1 (Fig.~\ref{fig-cmdplus}). Thus the bright cluster variables are not consistent with contamination by metal-poor intermediate age field stars or blue stragglers. Although we cannot come to a definite conclusion about their nature, it seems likely that they are indeed photometric blends.

\begin{deluxetable}{lc}
\tablecolumns{2}
\tablewidth{0pt} 
\tablecaption{Properties of DDO216-A1 \label{tab:props}}
\tablehead{
\colhead{Parameter} & 
\colhead{Value}}
\startdata
Right Ascension (J2000.0) & 23:28:26.3 \\
Declination (J2000.0)         & $+$14:44:25.2 \\
Projected dist.\ from galaxy center      & 6$\arcsec$, 26~pc \\
m814W (mag)                    & 18.64 $\pm$0.07\\
m814W$-$m475W (color)  &  1.47 $\pm$0.16\\
Distance modulus, m$-$M$_0$ & 24.77 $\pm$0.08 \\
Reddening, E(B-V)             & 0.16 $\pm$0.02 \\
M814W (mag)                     & $-$7.90 $\pm$0.16 \\
(M475W$-$M814W)$_0$      & 1.18 $\pm$0.16 \\
Central surface brightness (m$_{814}$ arcsec$^{-2}$) & 20.85 $\pm$0.17   \\
Half-light radius, r$_{\mathrm{h}}$   & 3$\farcs$1, 13.4~pc \\
Core radius, r$_{c}$ (arcsec)                  & 2.1 $\pm$0.9 \\
Concentration index (log(r$_{t}$/r$_{c}$) & 1.24 $\pm$0.39\\
Metallicity ([M/H])                & $-$1.6 $\pm$0.2 \\
Age (Gyr)                             & 12.3 $\pm$0.8 \\
log(M/M$_{\odot}$)   & 5.0 $\pm$0.1 \\
\# RR~Lyrae stars                &  $\approx$30 \\
Mean RRab period (d)                    &  0.556 $\pm$0.039\\
Mean RRab magnitude (V)               & 25.70 $\pm$0.11
\enddata
\end{deluxetable}

\section{Discussion and Summary} \label{sec:summ}

\subsection{Rediscovery of a Cluster}

The photometric and structural parameters of DDO216-A1 are summarised in Table~\ref{tab:props}. By every available measure, DDO216-A1 is a bona fide, ancient globular cluster, indistinguishable in many ways from the globular clusters of the Milky Way and Large Magellanic Cloud. While the cluster was first identified over 35 years ago, it has only sporadically been recorded in the literature \citep{gal98}. With an absolute magnitude M$_{\mathrm V}$ $\approx$ $-$7.1, DDO216-A1 contributes roughly 0.5\% of the V-band light of PegDIG-- this makes its lack of study all the more remarkable, given the long history of observational studies of PegDIG. 

The reasons for the omission of DDO216-A1 from lists of globular clusters in dwarf galaxies probably stem from a combination of the location of DDO216-A1, seen in projection against the densest part of PegDIG; the unusually extended nature of the cluster, which both gives it much less contrast with its surroundings and makes it somewhat unlike Galactic globular clusters; and the prevalence of background galaxies in the field. At a distance of 900~kpc, it requires diffraction limited imaging to resolve the cluster even to the level of the horizontal branch. The cluster is not prominently visible in Spitzer space telesope IRAC images \citep{jac06,boy09} due to its lack of asymptotic giant branch stars; with hindsight, it is obvious in SDSS images, although unresolved.

\subsection{Cluster Mass}

In the absence of spectroscopic information, the cluster mass can only be estimated in a model-dependent way or by comparison to better-studied, similarly luminous clusters. \citet{mcl05} compared the dynamical mass-to-light ratios for a large set of Milky Way and Magellanic Cloud globular clusters to predictions of population synthesis models by \citet{bru03}. Using their preferred initial mass function \citep{cha03}, they found a mean model M/L $\approx$1.9 for the old clusters. The median dynamical mass to light ratio for the clusters in their sample was 82$\pm$7\% of this value, with a substantial scatter. 

Our absolute magnitude for DDO216-A1 is M$_V$ = $-$7.14 $\pm$0.16, which translates to a V-band luminosity of (5.97 $\pm$0.95)$\times10^4$~L$_{\odot}$. Using the M/L estimates from \citet{mcl05} gives either 1.13$\pm$0.18 (population synthesis) or 0.93$\pm$0.15 (dynamical) $\times$10$^5$~M$_{\odot}$. The true range of possible values is even larger, because of potential variations in the initial mass function and the observed variations between clusters. Given the lack of kinematic constraints, an appropriate way to express the probable mass of the cluster is $\log$(M/M$_{\odot}$) = 5.0$\pm$0.1.  This is entirely consistent with mass estimates for similarly bright and extended, old clusters in the Milky Way \citep[e.g., IC4499,][]{han11}, and the LMC \citep[e.g., Reticulum,][]{sun92}.

\subsection{Formation, Migration, and Survival}

The cluster's projected position near the center of PegDIG raises the question of its provenance and survival. If the cluster formed in situ at the center of the galaxy, then it is natural to ask how it has survived tidal evaporation for 12~Gyr. Alternatively, if DDO216-A1 formed at an arbitrary location in the galaxy, then dynamical friction must be acting efficiently enough to bring it nearly to the center within its lifetime. 

Survivability of clusters in dwarf galaxies can be calculated probabilistically using analytical models for dynamical friction and cluster evolution (Leaman et al., in preparation). Using the dynamical friction formula from \citet{pet16}, a range of galaxy mass profiles and cluster orbits can be tested to see if there are any plausible initial conditions conducive to cluster inspiral and survival. The half-light radius of PegDIG is $\approx$700~pc \citep{kir14} and its stellar mass is log(M$_*$) $\approx$10$^7$~M$_{\odot}$ \citep{mcc12}, but its mass profile is not extremely well known. To a first approximation it could be taken as similar to a scaled-down version of WLM, which is well-fit by a \citet{nav97} profile with virial mass M$_{\rm vir}$ = 10$^{10}$~M$_{\sun}$ and concentration parameter $c$ = 15 \citep{lea12}.

To reflect the uncertainties in the parameters, we ran 2000 trials in which the important unconstrained parameters were drawn at random, and the dynamical friction and tidal destruction timescales were calculated analytically. The parameters and their range of sampled values were the initial distance and orbital eccentricity for DDO216-A1 (evenly distributed from 0--2~kpc and from 0--1, respectively), and the virial mass (log-normal distributed around 10$^{10}$~M$_{\odot}$), concentration index (normally distributed around $c$ = 12.5), and Einasto profile slope (evenly distributed between 0--1) for the PegDIG halo.

In this set of trials, 27\% of the clusters are found to have survival times longer than 12~Gyr and dynamical friction timescales shorter than this. Within the range of parameters considered, there were no strong trends of survivability in the concentration index, Einasto profile slope, or virial mass, but the best cluster survivability is found for birthplaces from $\approx$300-1000~pc from the galaxy center; nearly half of 10$^5$~M$_{\odot}$ clusters born within this range sink to within $\lesssim$100~pc of the center without tidal destruction over their lifetime. Clusters born interior to this region tend to be tidally disrupted, and clusters born in the galaxy outskirts have dynamical friction timescales longer than a Hubble time. The general feature of the analysis, that in many cases clusters will be disrupted, but that the most massive will sometimes survive to be observed near the center of the host galaxy, is consistent with advanced numerical simulations of star formation in dwarfs (Charlotte Christensen, private communication).

\citet*{gui16} performed hydrodynamical simulations of a larger dwarf (M$_*$ = 10$^{9.5}$~M$_{\odot}$), and observed exactly this behavior, producing a 10$^8$~M$_{\odot}$ nuclear star cluster as the result of inspiral, gas accretion, and merging of an initially 10$^4$~M$_{\odot}$ protocluster formed in the outskirts of the dwarf. Their simulated cluster arrives in the central part of the dwarf after $\approx$1~Gyr and is quenched by a final merger with another large cluster. This raises the possibility that DDO216-A1 might show an extended history of star formation as the result of dry or wet mergers, although there is little evidence for this in the current data.

These results show that the cluster location is consistent with formation across a large volume of PegDIG, excluding the central region where it is now observed. Given the propensity of clusters to dissolve when located at the center of the galaxy, it seems unlikely that DDO216-A1 formed at its current location. Because dynamical friction tends to stall when the cluster reaches the radius at which the host galaxy density profile flattens into a core, it is not surprising that the cluster is not observed at the precise center of PegDIG. Both of these factors point to the likelihood that the true distance from the PegDIG center to the cluster is larger than the projected separation.

\begin{figure*}[t!]
\plotone{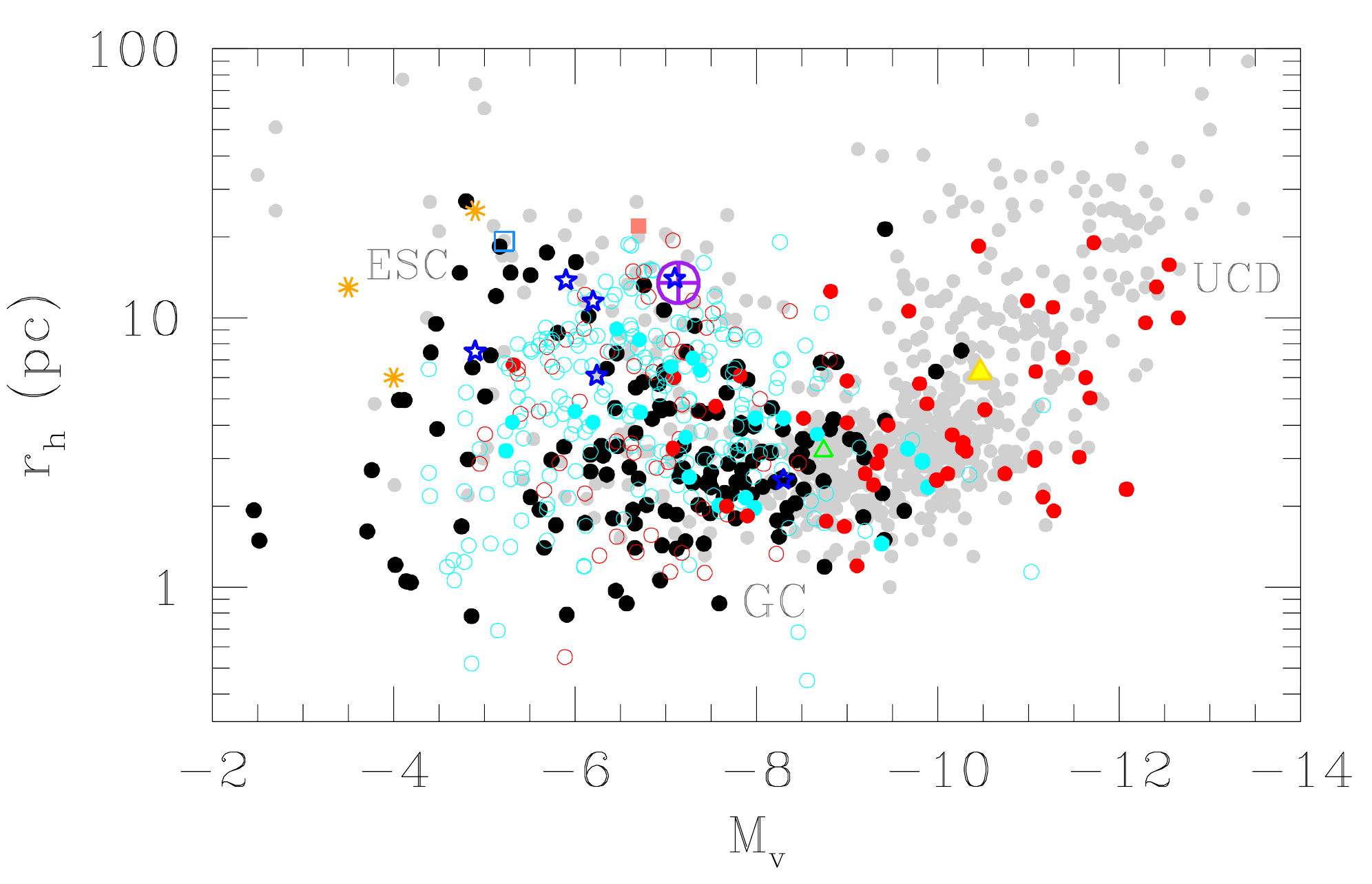}
\caption{Half-light radius and absolute magnitude for a selection of stellar systems. Following \citet{bro11}, the regions of the (r$_h$,M$_V$) plane are labelled ESC-- extended star clusters, UCD-- ultracompact dwarfs and nuclei, and GC--globular clusters. Legend: {\it purple circled cross:} DDO216-A1; {\it grey circles:} clusters and dwarf galaxies \citep{bro11}; {\it black circles:} Milky Way globulars \citep[][2010 edition]{har96}; {\it red circles:} clusters in early-type dwarfs. Filled circles denote galaxy nuclei and other clusters within 150~pc of the host center \citep{cot06,sha05,geo09a}; {\it cyan circles:} clusters in late-type dwarfs, from the same sources. Filled and open symbols as above; {\it orange asterisks:} clusters in Local Group dwarfs fainter than PegDIG; {\it blue stars:} clusters in NGC~6822; {\it green triangle:} WLM cluster; {\it pink filled square:} Scl-dE1-GC1; {\it light blue open square:} Reticulum; {\it gold triangle:} M54 (Sagittarius nucleus). See text for references to individual objects. 
\label{fig:rhmv}}
\end{figure*}

\subsection{DDO216-A1 in Context} 

Following \citet{bro11}, we show the half-light radius and absolute magnitude of a sample of stellar systems in Figure~\ref{fig:rhmv}. The grey circles are points from \citet{bro11}, and include extragalactic globular  (GC) and extended (ESC) clusters,  and many ultra-compact dwarf galaxies (UCD). The Milky Way globular clusters are plotted in black.  Additional clusters drawn from surveys of dwarf galaxies are plotted in red (early-type galaxies) and cyan (late-type galaxies); clusters identifed as nuclear \citep{cot06} or within projected distance 150~pc of their host centers are plotted as filled circles; the symbols for off-center clusters are open. DDO216-A1 clearly sits within the range of extended clusters in this plane, and appears characteristic of the clusters in late-type dwarf galaxies. Some specific clusters mentioned in the text are highlighted in Fig.~\ref{fig:rhmv} with alternate symbols.

A census of globular clusters in Local Group dwarf galaxies has been given in \citet{for00}; the count has updated slightly since then as the result of HST and wide-field surveys of dwarfs at large distances, at low surface brightness, or where foreground confusion is high. However, most of the additions are small clusters at least an order of magnitude less luminous than DDO216-A1\footnote{At least one cluster, the one listed in Table~1 of Forbes et al.\ (2000) for DDO~210, has been removed from the list after subsequent imaging.}. Notable exceptions to this trend are the brightest 4 of the 7 new clusters discovered in NGC~6822 by \citet{hwa11} and \citet{hux13}; NGC~6822-SC1 is of similar age, half-light radius, and luminosity to DDO216-A1, although it is significantly more metal-poor \citep{vel15}. 

Three Local Group galaxies fainter than PegDIG host star clusters: Eri~II \citep[M$_{\mathrm V}$ = $-$7.1,][]{crn16}, And~XXV \citep[M$_{\mathrm V}$ = $-$9.7,][]{cus16}, and And~I \citep[M$_{\mathrm V}$ = $-$11.7,][]{gre00}. Each hosts a single cluster, but all are far fainter than DDO216-A1 (M$_{\mathrm V}$ $\approx$ $-$3.5 to $-$5), and are quite diffuse and extended by comparison.  These galaxies are all dSphs;  the cluster DDO216-A1 is as luminous as the entire galaxy Eri~II. These clusters are plotted as orange asterisks in Fig.~\ref{fig:rhmv}.

The cluster-hosting dIrr galaxies NGC~6822 (M$_{\mathrm V}$ = $-$15.2) and WLM (M$_{\mathrm V}$ = $-$14.2) are both far more luminous than PegDIG. The lowest-luminosity Local Group galaxies with recorded globular clusters as bright as DDO216-A1 are the Fornax and Sagittarius dwarf spheroidals \citep{mcc12}. Both galaxies are $\approx$1~mag brighter than PegDIG and each has multiple clusters. Fornax hosts 5 globulars, and Sagittarius at least 4 \citep{dac95}, possibly as many as 9 \citep{law10}. Normalising each galaxy to an absolute magnitude of M$_{\mathrm V}$ = $-$15, the specific frequency of globular clusters is 29 for Fornax, 5--9 for Sagittarius, 7 for NGC~6822, 2 for WLM, and 10 for PegDIG.  PegDIG thus has a rather ordinary specific cluster frequency compared to other Local Group galaxies and its role as cluster host is not surprising. 

Statistics of clusters in dwarf galaxies beyond the Local Group are necessarily less complete, but summaries of statistical properties of clusters in dwarfs can be found in, e.g., \citet{sha05,geo09a,zar16} and references therein. DDO216-A1 appears to be a typical old and metal-poor cluster of the type common to both giant and dwarf galaxies, although it is unusual to find a cluster as luminous as DDO216-A1 in a galaxy as faint as PegDIG. The contribution of the cluster to the total light of the galaxy is $\approx$0.5\%;
\citet{lar15} has shown that globular clusters in dwarf galaxies can contribute up to 20--25\% of the metal-poor stars in a given dwarf. While a detailed analysis awaits the full SFH of PegDIG, it appears as though the fraction is much smaller in this case.
Unlike  Fornax, WLM, and NGC~6822, the cluster does not have a dramatically lower metallicity than the galaxy as a whole, only about 0.3~dex less. 

The structural parameters of the cluster resemble those of the ``faint fuzzy'' clusters found in lenticular galaxies \citep{bro02}, but its color is much bluer and it otherwise appears typical of the globular cluster population of dwarfs. DDO216-A1 is unusually close to the center of its host galaxy, and is also unusually extended for a cluster with small projected galacticentric distance \citep{sha05}. However, it falls below the luminosity and surface brightnesss of the great majority of clusters that are most likely to be identified as galactic nuclei \citep{geo09b,bro11}, although exceptions exist \citep{cot06,geo14}. \citet{sha05} find that among dSphs with globular clusters, more than half show clusters seen in projection against the center of the galaxy. While DDO216-A1 fits among these clusters in luminosity, its large radius distinguishes it from the (usually) compact central clusters. However, it is still quite a bit more compact than very extended clusters like Scl-dE1~GC1, with a half-light radius of 22~pc \citep{dac09}. 

PegDIG is the lowest-mass gas-rich galaxy in the Local Group with a major star cluster. PegDIG is a transition type galaxy, with characteristics of both irregular and spheroidal galaxies; its principal dSph-like quality is the very low rate of current star formation \citep{ski97}. In the (r$_h$, M$_{\mathrm V}$) plane, DDO216-A1 is more typical of clusters in dSph galaxies than in dIrr \citep{sha05}. If PegDIG is considered to be a dIrr, then DDO216-A1 would be one of the largest clusters known in a small dIrr. More extreme examples are very rare. For example, in the list of \citet{geo09a}, there is only one less-luminous dIrr with a comparably bright cluster, the M$_{\mathrm V}$ = $-$11.9 field galaxy D634-03, at a distance of 9.5~Mpc.  By comparison, there are numerous cases of globulars in dSphs at the same host galaxy absolute magnitude. The overall trend with galaxy morphology appears to suggest that globular cluster populations are typically much poorer among the gas-rich dwarfs, which also lack nuclear clusters. 

DDO216-A1 is more extended than $\approx$90\% of Galactic and Magellanic Cloud globular clusters, although given its absolute magnitude and half-light radius it is not an extreme outlier. It would not be out of place among ``outer halo'' globulars such as IC~4499 or NGC~5053, Magellanic Cloud clusters like Reticulum (LMC), or clusters like NGC~6822-SC1.  Contrastingly, it is much more luminous than the extended clusters Fornax 1, Arp 2, Ter 8, or Pal 12, similarly extended clusters that are either definite or probable members of spheroidal galaxies. 

Given the complexity of the field star background, it is impossible to say whether there are multiple ages or metallicities present in the cluster without spectroscopy. These features would be indicative of cluster mergers or in situ star formation from newly accreted gas in the cluster, either of which could contribute to the formation of a nuclear star cluster \citep[e.g.,][]{gui16}. However, circumstantial evidence argues against it being a nuclear cluster: it would be an outlier from the host mass-cluster radius and -cluster mass relations presented in \citet{geo16}, being both too extended for its mass and slightly too massive for its host compared to other nuclear star clusters.

\subsection{Summary}

We have imaged the central, extended star cluster in the Pegasus transition dwarf with HST/ACS, and obtained a color-magnitude diagram reaching $\approx$0.5~mag above the cluster main-sequence turnoff. DDO216-A1 is in some respects a typical globular cluster but is more extended than most. We find in particular the following major features:

\begin{enumerate}
\item We have confirmed that DDO216-A1 is a bona fide globular cluster, with absolute I-magnitude M$_{814}$ = $-7.90 \pm0.16$, a mass of $\sim$10$^5$~M$_{\odot}$, and a half-light radius $r_h \approx 13$~pc. While it is larger than $\approx$90\% of Milky Way globular clusters, it is structurally similar to some of the outer halo Milky Way globular clusters, and is not an extreme example of the type of extended cluster that seems to be characteristic of some dwarf galaxies. 

\item Based on the color-magnitude diagram, the cluster is ancient, 12.3 $\pm$0.8~Gyr old, and is moderately metal-poor, [Fe/H] = $-$1.6 $\pm$0.2. In its variable star properties, DDO216-A1 harbors $\approx$30 RR~Lyrae stars, and is an Oosterhoff Type I cluster with a specific frequency S $\approx$42.

\item DDO216-A1 contributes $\approx$0.5\% of the V-band luminosity of PegDIG, but the galaxy does not have an anamolously high specific frequency of clusters compared to other dwarf galaxies. Despite the low mass of the host, the cluster is very close to the peak of the globular cluster luminosity function for all Local Group galaxies.

\item The cluster is seen in projection within 30~pc of the galaxy center, but it is much more extended, for its mass, than the typical nuclear star cluster \citep[e.g.][Figure 3]{geo16}. We have not detected evidence for multiple stellar populations in the cluster. Birth at a distance of $\sim$0.3--1~kpc and subsequent infall due to dynamical friction is a possible scenario resulting in the observed cluster position.  Because dynamical friction is inefficient within cored mass distributions, and the cluster has not been tidally disrupted, it is likely that the true distance between the cluster and galaxy center is a few times larger than the projected separation. 
\end{enumerate}

The association of globular clusters with intense episodes of star formation involving very large gas masses \citep{bro06} indicates that PegDIG might have had a tumultuous early history.  The observed relationship between the size of the largest star cluster and the peak SFR \citep[e.g.,][]{lar02,coo12} suggests that PegDIG should have experienced its highest SFR around the time that DDO216-A1 formed. Considering the large star formation surface densities associated with the formation of a cluster as massive as DDO216-A1 \citep[e.g.,][]{joh16} and the constraints on the total stellar mass of the galaxy, this raises the likelihood that PegDIG may have formed a large fraction of its stars in an intense burst around the time of cluster formation. The resulting peak in SFR at early times would tend to make PegDIG more similar to a prototypical dSph than to a dIrr.

Although PegDIG is not currently close to any other known system, its radial velocity and distance from M31 suggest that it is not likely to be on its first infall into the Local Group. \citet{gar14} have found in their cosmological simulations of structure formation that in mock Local Groups, the majority of dwarf galaxy-sized subhaloes found within 1--1.5 virial radii of a large halo at redshift $z$ = 0 have previously spent time inside the virial radius. Timing arguments using numerical action reconstructions \citep[e.g.,][]{sha13}, while subject to significant uncertainty due to unknown initial conditions, also suggest that PegDIG has not always been isolated.

Close encounters with M31 or another dwarf could have dramatically increased the SFR and therefore the statistical probability for a large cluster to be formed. \citet{dea14} predict from cosmological simulations that most dwarf-dwarf major mergers tend to occur near the time of first infall into the virial radius of a larger parent galaxy, consistent with the large age of DDO216-A1. The next paper in this series will examine the complete star formation history of PegDIG; the sample CMD for an outer field indicates that photometry well below the oldest MSTO will allow a detailed reconstruction of the SFH back to the oldest ages. 

\acknowledgments 
Support for this work was provided by NASA through grant number HST~GO-13768 from the Space Telescope Science Institute, which is operated by AURA, Inc., under NASA contract NAS5-26555. MBK acknowledges support from NSF grant AST-1517226 and from NASA grants HST-AR-12836, HST-AR-13888, and HST-AR-13896 awarded by STScI. The authors thank the anonymous referee for their very helpful comments, which improved the paper. Thanks to Charlotte Christensen for sharing her simluation work on the formation of dwarf galaxies. AAC is grateful to Chris Howk, Jay Gallagher, and Warren Hankey for help chasing obscure references. This research made extensive use of NASA's Astrophysics Data System bibliographic services.

\clearpage

\end{document}